\documentclass[%
 reprint,
 amsmath,amssymb,
 aps,
]{revtex4-1}

\usepackage{units}
\usepackage{graphicx}
\usepackage{dcolumn}
\usepackage{bm}
\usepackage{color}

\begin{document}

\preprint{APS/123-QED}

\title{Electronic transport in high magnetic fields of thin film MnSi}

\author{Nico Steinki$^{1}$, David Schroeter$^{1}$, Niels W\"achter$^{1}$, Dirk Menzel$^{1}$, Hans Werner Schumacher$^{2}$, Ilya Sheikin$^{3}$ and Stefan S\"ullow$^{1}$}

\affiliation{
$^{1}$Institut f\"ur Physik der Kondensierten Materie, Technische Universit\"at Braunschweig, D-38106 Braunschweig, Germany\\
$^{2}$Physikalisch-Technische Bundesanstalt, D-38116 Braunschweig, Germany\\
$^{3}$Laboratoire National des Champs Magn$\acute{\textrm{e}}$tiques Intenses (LNCMI-EMFL), CNRS, UGA, F-38042 Grenoble, France}

\date{\today}

\begin{abstract}
We present a study of the magnetoresistivity of thin film MnSi in high magnetic fields. We establish that the magnetoresistivity can be understood in terms of spin fluctuation theory, allowing us to compare our data to studies of bulk material. Despite of a close qualitative resemblance of bulk and thin film data, there are clear quantitative differences. We propose that these reflect a difference of the spin fluctuation spectra in thin film and bulk material MnSi.
\end{abstract}


\maketitle

\section{Introduction}

The cubic helimagnet MnSi has intrigued researchers in the field of solid state magnetism for over half a century \cite{williams1966,shinoda1966,pfleiderer2010}. The material, belonging to the class of $B20$ compounds, was originally studied in the context of spin fluctuation theory \cite{sakakibara1982,moriya1985}. Later, the pressure induced suppression of helical magnetic order ($T_N = 29$\,K at ambient pressure) became the focus of studies in the context of quantum criticality in itinerant $d$-metals \cite{pfleiderer2010,thompson1989}. Finally, it was noted that the early-reported field-induced {\it A-}phase in MnSi \cite{sakakibara1982} does represent a skyrmion lattice phase \cite{muehlbauer2009}, this way establishing the material as a model compound for studies of skyrmion physics.

Especially in the latter context of skyrmionics, in recent years various efforts have been undertaken to grow MnSi thin films. Conceptually, the idea is based on the notion that skyrmionic phases should be energetically favored in two-dimensional structures. Hence, in order to study skyrmionic properties in solid state materials, thin film MnSi has been a prominent candidate to perform corresponding studies \cite{karhu2010,karhu2011,karhu2012,geisler2012,wilson2012,engelke2012,li2013,suzuki2013,menzel2013,wilson2013,yokouchi2014,wilson2014,engelke2014,meynell2014a,meynell2014b,yokouchi2015,lancaster2016,figueroa2016,meynell2016,meynell2017,trabel2017,schroeter2018}. Surprisingly, while these studies brought various insights into the relationship of the properties of thin film MnSi and the corresponding bulk behavior, there are also a few quite striking differences.

First of all, in comparison to bulk material, thin film MnSi undergoes a transition into a helical magnetic phase below $T_N \sim 45$\,K \cite{karhu2010,karhu2011,karhu2012,geisler2012,wilson2012,engelke2012,li2013,suzuki2013,menzel2013,wilson2013,yokouchi2014,wilson2014,engelke2014,meynell2014a,meynell2014b,yokouchi2015,lancaster2016,figueroa2016,meynell2016,meynell2017,trabel2017,schroeter2018}. This enhanced $T_N$ is attributed to the tensile strain in the MnSi film, exerted by the mismatch of the lattice parameters of MnSi and the underlying Si substrate. Effectively, it leads to a state of negative pressure in thin film material, with the bulk $T_N$ recovered if the films are pressurized \cite{engelke2014}.

Moreover, the structural anisotropy induced by the tensile strain affects the in-field properties, leading for instance to an increased critical field $B_C$ into the magnetically polarized state \cite{menzel2013}. Most strikingly, as yet there is no direct experimental evidence for a skyrmion lattice phase for the so-called out-of-plane geometry in thin film MnSi. As well, there is no final consensus if this ''non-observation'' of a skyrmion lattice is an intrinsic or extrinsic property. On the one hand, it was argued that the effective negative pressure and pressure induced anisotropy drives thin film MnSi into a parameter range where a skyrmion lattice would not be stable anymore \cite{karhu2012}. On the other hand, the thin films - even if grown epitaxially - arise from island growth on the Si substrate. This results in merohedrally twinned thin films, {\it i.e.}, the films contain left- and right-handed crystallites. In this situation, the corresponding skyrmions would have opposite sense of rotation, and would annihilate upon meeting. Here, of course, the lack of observation of a skyrmion lattice would be extrinsic, as it results from the non-mono-chiral character of the films.

The difference in behavior of thin film and bulk MnSi begs the question if they can be related to fundamental material properties. Bulk MnSi is a prime example of a system, where spin fluctuation theory has been invoked to quantitatively describe the material properties. Here, an analogous study of thin film MnSi seems worthwhile, with issues such as the influence of residual disorder in the films, structural low- (two)-dimensionality or the effectively negative pressure possibly becoming relevant. Therefore, we have set out to perform a high field magnetoresistivity study of thin film MnSi. In our approach, we closely follow in procedure and compare our data to a seminal study of the magnetoresistive properties of bulk single crystalline MnSi \cite{sakakibara1982}. Based on the comparison, we discuss the electronic properties of thin film MnSi in terms of spin fluctuation theory.   

\section{Experimental}

For the high magnetic field studies, two two epitaxial grown MnSi thin films (thickness sample \#1: 10 nm; sample \#2: 30 nm) were synthesized by molecular beam epitaxy on [111] Si-substrate as described previously \cite{schroeter2018}. To enable a direct comparison between MnSi thin films and MnSi bulk material regarding the magnetotransport properties the MnSi thin film samples were micro-structured by electron beam lithography (see Ref. \cite{schroeter2018}). This allows resistance measurements with a conventional four point AC-method in the same geometry as for bulk material. The structures are 50 $\mu$m wide and the voltage leads have a distance of 70 $\mu$m. 

The high magnetic field measurements were performed on the thin film samples at the Laboratoire National des Champs Magn$\acute{\textrm{e}}$tiques Intenses in Grenoble, France. The magnetoresistivity was measured at various temperatures in a range between 3 and 100 K in an external magnetic field \textit{B} up to 24 T. For both samples the external magnetic field was applied perpendicular and parallel to the sample current \textit{I} ($B$\,$\perp$\,$I$ and $B$\,$\parallel$\,$I$). In the former configuration, the field was applied out-of-plane, in the latter in-plane.

For direct comparison to the results from Ref. \cite{sakakibara1982}, we also studied MnSi bulk crystals synthesized by the Czochralski tri-arc method. The single crystals have been characterized with respect to their essential physical properties as described below. The transverse magnetoresistivity ($B$\,$\perp$\,$I$) of the bulk samples was determined in a temperature range between 2 and 100 K in an external magnetic field up to 8 T. In addition the zero field resistivity of all samples was measured in a temperature range between 2 and 300 K.   

\section{RESULTS}

\begin{figure}
\centering
		\includegraphics[width=1 \columnwidth]{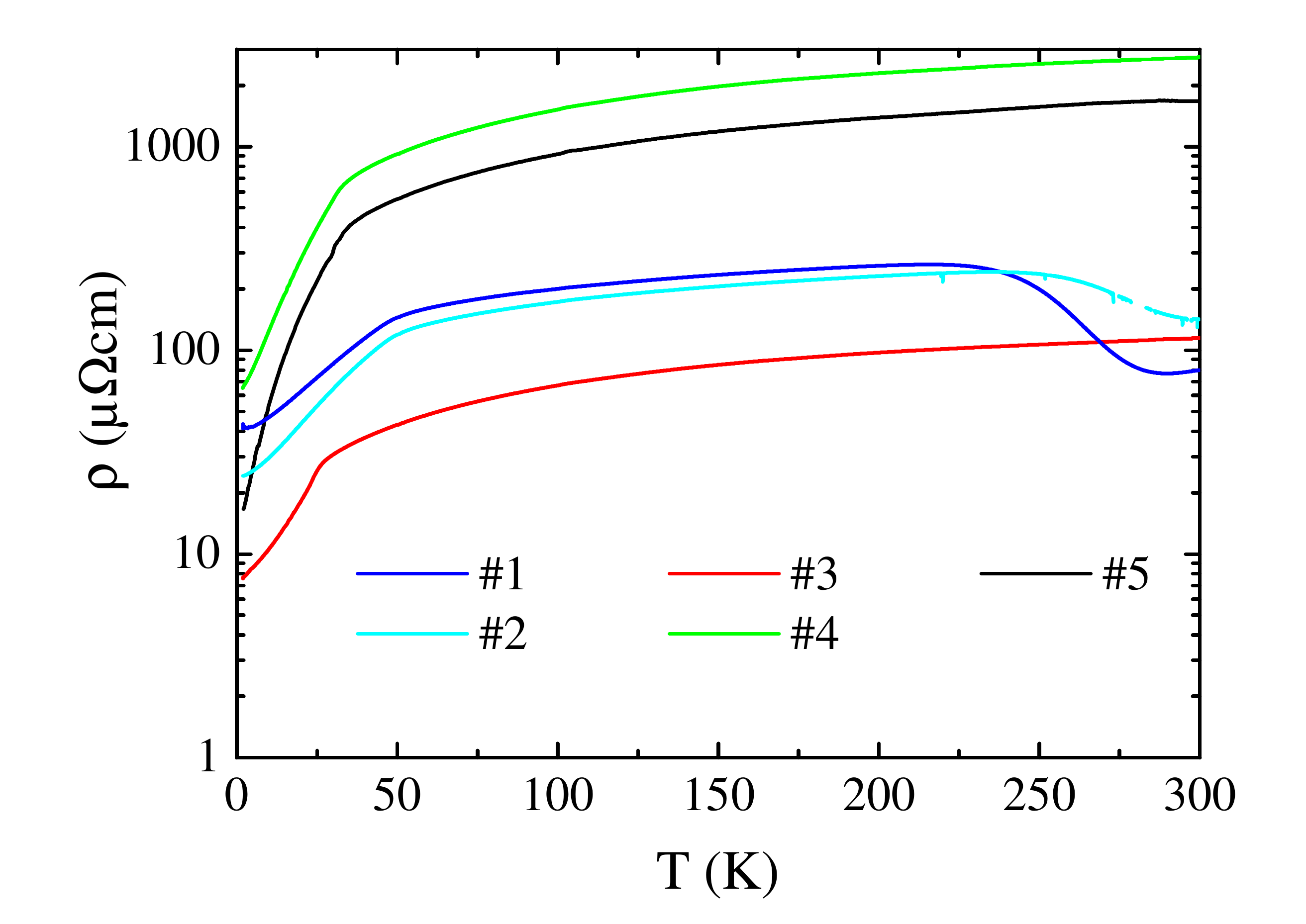}
	\caption{(Color online) Zero field resistivity of different samples MnSi (\#1: thin film 10 nm thickness, \#2: thin film 30 nm thickness, \#3: bulk with RRR of 15, \#4: bulk with RRR of 42, \#5: bulk with RRR of 104); for details see text.}
	\label{fig1}
\end{figure}

In Fig. \ref{fig1} we plot the zero field resistivities of our bulk and thin film samples. Qualitatively, the overall behavior is similar to that reported previously for single-crystalline and thin film material. For both types of systems, overall there is a metallic temperature dependence, with an anomaly at the transition at $T_N$ into the helical state, a $T^2$-like behavior below $T_N$ and some rounding of the resistivity above $T_N$ attributed to spin fluctuations on top of the phononic resistivity. In detail, however, there are a few issues to be noted.

First, while all single crystals MnSi have the same transition temperature into the helimagnetic state at $T_N = 29$\,K (which was also verified by susceptibility measurements, not shown) and exhibit a similar temperature dependence of the resistivity $\rho$, absolute values of $\rho$ vary by more than an order of magnitude. It suggests that the determination of the absolute value of $\rho$ is affected by cracks in the single crystalline samples. Then, in terms of characterizing the crystalline quality of our bulk samples, instead of the residual resistivity $\rho_0$ the best measure is the residual resistivity ratio (RRR) here defined as $\rho_{300 \rm K}/\rho_{2 \rm K}$. We find values RRR for sample \#3: RRR = 15, \#4: RRR = 42 and \#5: RRR = 104. We thus conclude that sample \#5 has the highest crystalline quality, even though the room temperature resistivity is nonphysically high with 1700\,$\mu \Omega$cm. For comparison, from the experimental data published in Ref. \cite{sakakibara1982} we estimate a value RRR for the crystal studied in that work of the order of 60. Nowadays, for MnSi, in general a RRR of the order of 100 is taken to signal a ''good sample quality'' \cite{adams2011}.

Second, for the thin film samples we find transition temperatures into the helimagnetic state at $T_N = 45$\,K, consistent with previous reports \cite{karhu2010,karhu2011,karhu2012,geisler2012,wilson2012,engelke2012,li2013,suzuki2013,menzel2013,wilson2013,yokouchi2014,wilson2014,engelke2014,meynell2014a,meynell2014b,yokouchi2015,lancaster2016,figueroa2016,meynell2016,meynell2017,trabel2017,schroeter2018}. Only, at high temperatures both film samples exhibit a downturn of $\rho$, different from the single crystal behavior. As we have demonstrated in Ref. \cite{schroeter2018}, the downturn arises from a breakdown of a Schottky barrier between thin film MnSi and the Si substrate. In result, at high temperatures the MnSi film is shortcut by the substrate, leading to the downturn in $\rho$. We note that - while we need to keep aware of these experimental artifacts - they will not affect the magnetoresistive behavior reported below, as we can use normalized representations of the magnetoresistive behavior and we discuss only low-temperature data not affected by the shortcut.

\begin{figure}
\centering
		\includegraphics[width=1 \columnwidth]{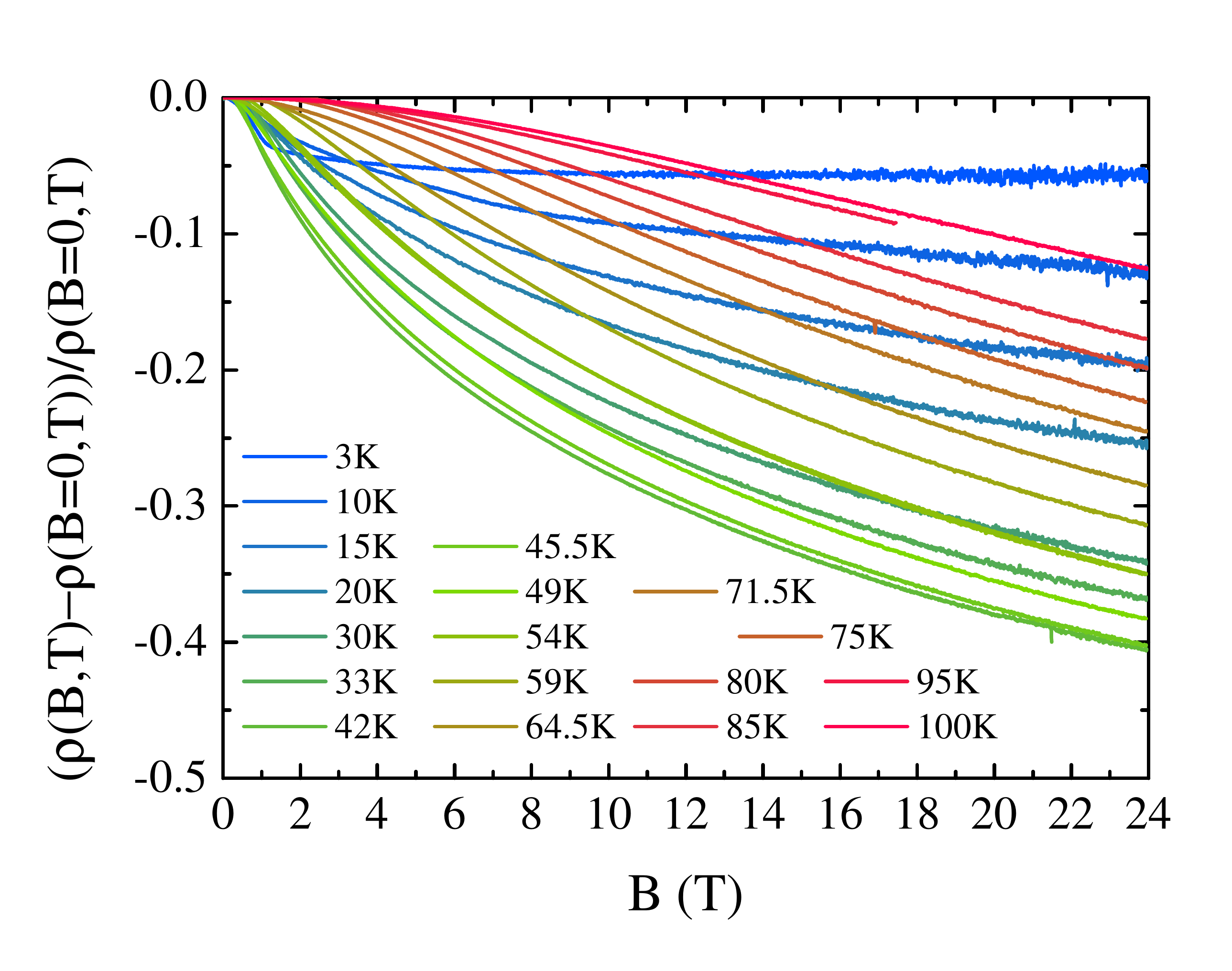}
	\caption{(Color online) Magnetoresistivity in transverse magnetic fields up to 24\,T for a 30\,nm thin film MnSi; for details see text.}
	\label{fig2}
\end{figure}

Subsequently, we have carried out an extensive run of magnetoresistivity measurements. As an example, in Fig. \ref{fig2} we plot the magnetoresistivity $MR$ defined as $MR = \left( \rho (B,T) - \rho (B = 0, T) \right) / \rho (B = 0, T)$ in transverse magnetic fields $B$ up to 24\,T at temperatures between 3 and 100\,K for the 30\,nm thick sample MnSi. Qualitatively and semiquantitatively, as will be documented below, the general behavior reported here for the 30\,nm sample in transverse geometry is similarly seen for the 10\,nm sample and the second field alignment.

For low temperatures, the magnetoresistivity exhibits an inverted $S$-shaped behavior which is easily saturated in a few T. As temperature is increased up to $T_N$, the $MR$ increases as well and the inverted $S$-shaped behavior broadens without reaching saturation. Finally, above $T_N$ only the downward curvature of the $MR \propto B^2$ remains from the inverted $S$-shaped character, with the overall size of the $MR$ being reduced again relative to the signal close to $T_N$ in the field range covered.

\begin{figure}
\centering
		\includegraphics[width=1 \columnwidth]{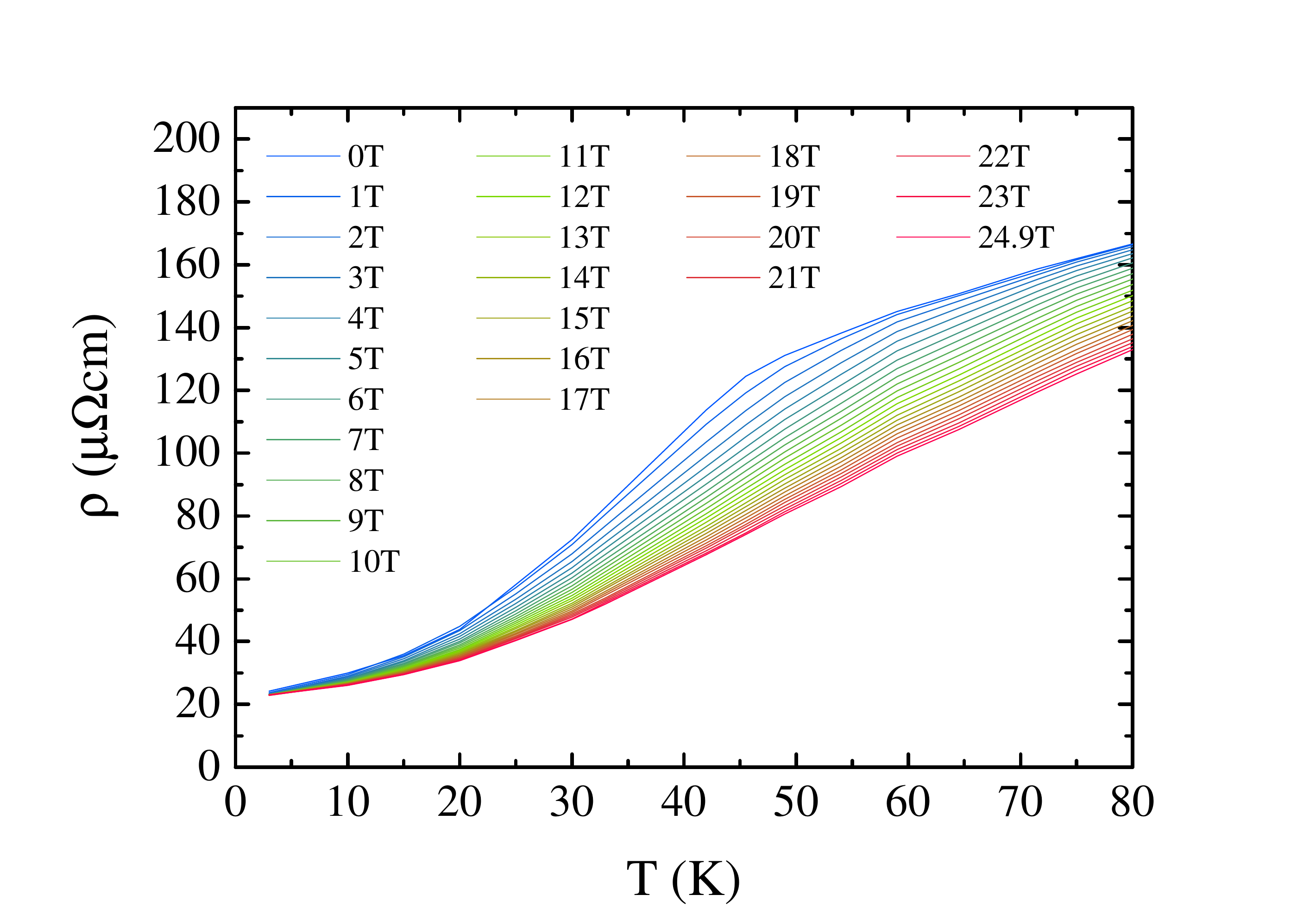}
	\caption{(Color online) Resistivity in transverse magnetic fields up to 24\,T for a 30\,nm thin film MnSi; for details see text.}
	\label{fig3}
\end{figure}

From the data presented in Fig. \ref{fig2}, for comparison with Ref. \cite{sakakibara1982} we construct the temperature dependence of $\rho$ in transverse magnetic fields up to 24\,T for the 30\,nm thick sample MnSi, which we plot in Fig. \ref{fig3}. Qualitatively, the behavior is similar to that reported in Ref. \cite{sakakibara1982} (see Fig. 6 therein), with a suppression of the resistivity in magnetic fields over a wide temperature range and the largest effect close to $T_N$. We note that the critical field of the helical phase of MnSi is about 1\,T, accounting for the observed disappearance of the kink in the resistivity in the magnetic fields plotted here. Of course, the qualitative similarity in behavior of single crystal and thin film material holds as well for the measurements of the 10\,nm sample and the second field alignment (not shown).

\begin{figure}
\centering
		\includegraphics[width=1 \columnwidth]{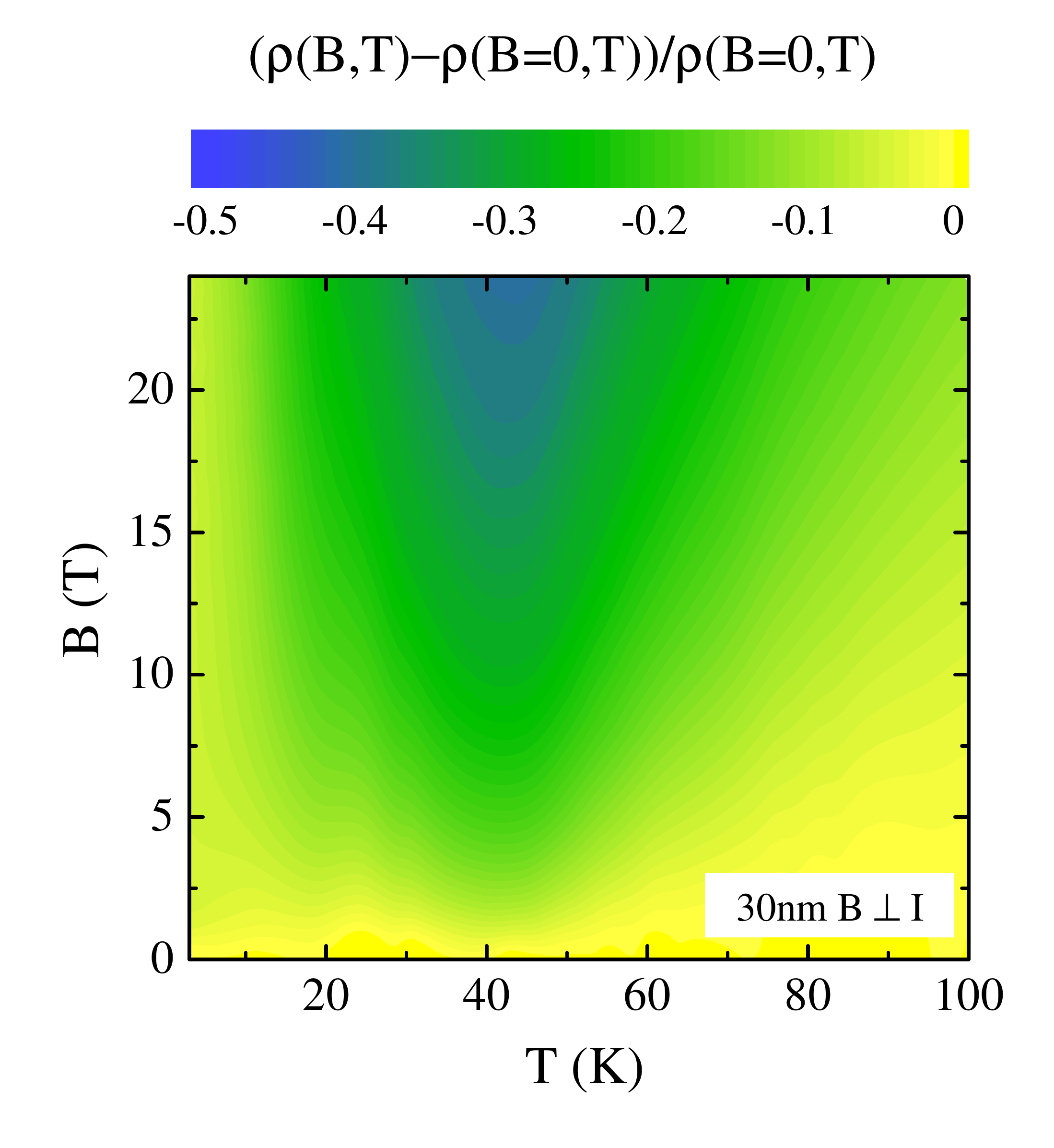}
	\caption{(Color online) Contour plot of the magnetoresistivity $MR$ in transverse magnetic fields up to 24\,T for a 30\,nm thin film MnSi; for details see text.}
	\label{fig4}
\end{figure}

The interpretation of the magnetoresistivity of MnSi has invoked the suppression of spin fluctuations in magnetic fields. These spin fluctuations are present over a very wide temperature range. A different approach to illustrate this effect is to present the data in form of contour plots, as we do in Fig. \ref{fig4} for the data from Fig. \ref{fig2}. In this representation, the color coding reflects the size of the magnetoresistivity, which nicely illustrates the most pronounced negative magnetoresistivity to occur close to $T_N$. 

Interestingly, the contour plot demonstrates that there is some asymmetry in the magnetoresistivity. The magnetoresistivity is somewhat stronger at temperatures $T < T_N$ than at $T > T_N$. Within a view of the $MR$ as just reflecting the suppression of spin fluctuations, one would naively argue that magnetic order already removes some of the spin fluctuations, as it is indicated by the downturn of the zero field resistivity at $T_N$. Then, for a normalized quantity as the $MR$ this should show up as a comparatively smaller signal, if compared to a situation were no spin fluctuations have been removed in zero magnetic field. Therefore, the asymmetry would be expected to appear as a stronger $MR$ above $T_N$, and as it has been shown for instance for the itinerant weak ferromagnet NbFe$_2$ \cite{steinki2018}. 

\begin{figure}
\centering
		\includegraphics[width=1 \columnwidth]{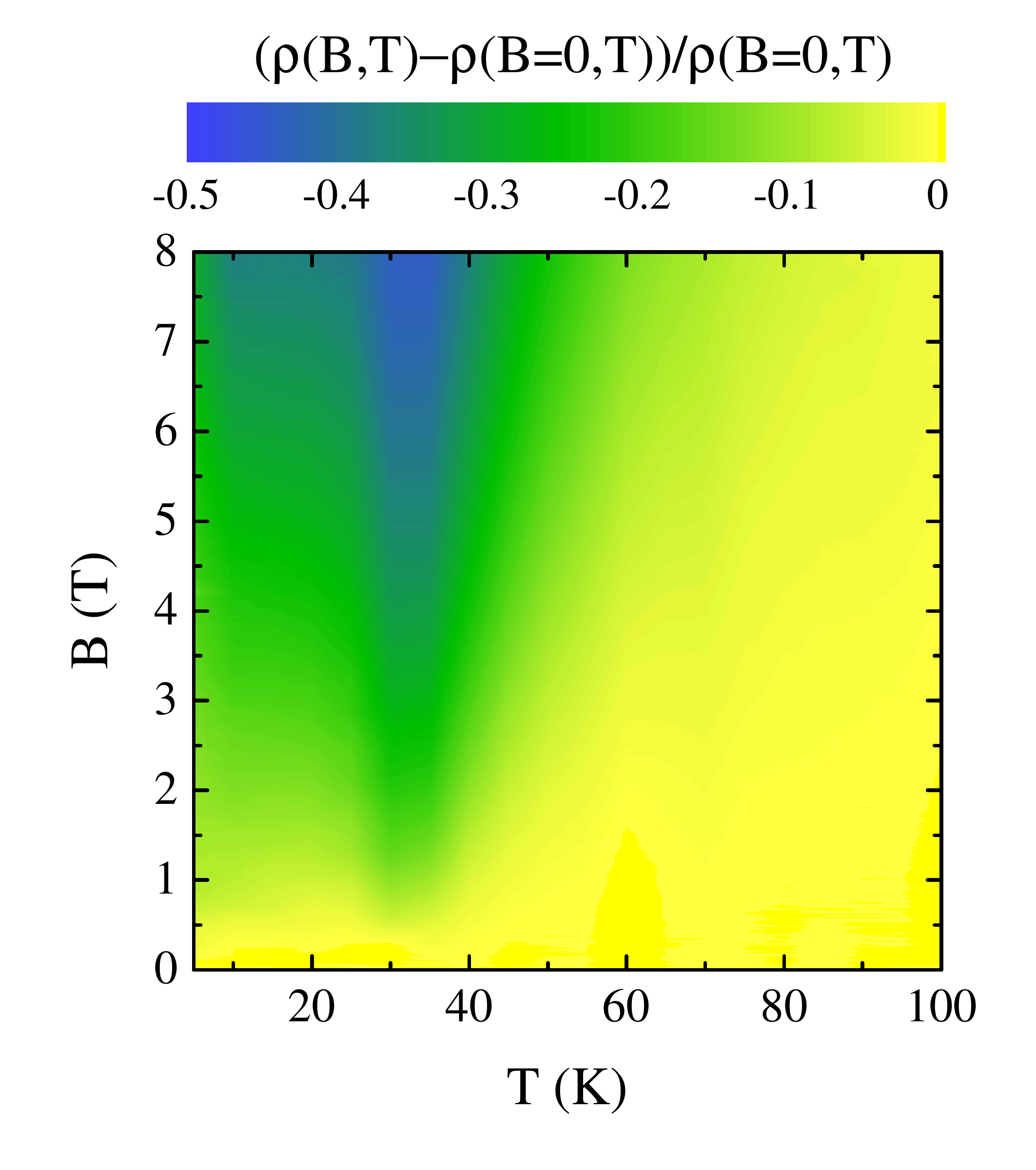}
	\caption{(Color online) Contour plot of the magnetoresistivity $MR$ in transverse magnetic fields up to 8\,T for a single crystal MnSi, sample \#5; for details see text.}
	\label{fig5}
\end{figure}

In fact, corresponding magnetoresistivity measurements on single crystalline MnSi (now up to only 8\,T) seem to be more in line with the observations made for single crystalline NbFe$_2$. This is illustrated in Fig. \ref{fig5}, where we display a contour plot of the $MR$ of sample \#5 in the same fashion as the thin film data. Here, the asymmetry in the magnetoresistivity is clearly tilted towards temperatures $T > T_N$, reflecting a stronger suppression of spin fluctuations in the paramagnetic phase. Altogether, as a qualitative interpretation, while the overall magnetoresistive behavior of bulk and thin film material exhibits a strong resemblance, there seem to be residual subtle differences in some aspects. If these subtle differences are extrinsic or intrinsic is not entirely clear. It seems conceivable that a distribution of transition temperatures in the films causes a smearing of the magnetoresistive features towards lower temperatures, and thus would be the result of a somewhat larger structural inhomogeneity in films than in single crystals.

\begin{figure}
\centering
		\includegraphics[width=1 \columnwidth]{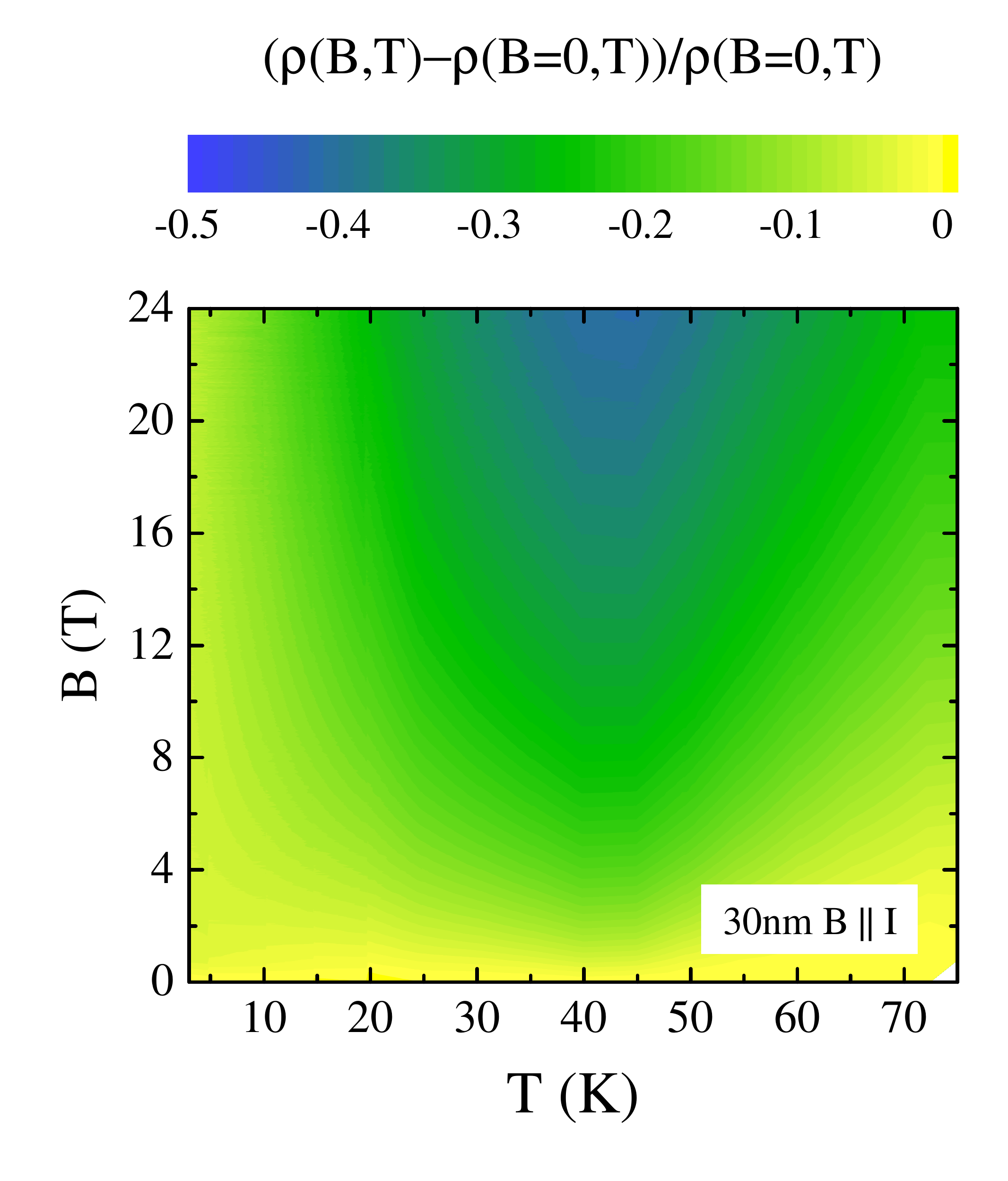}
	\caption{(Color online) Contour plot of the magnetoresistivity $MR$ in parallel magnetic fields up to 24\,T for a 30\,nm thin film MnSi; for details see text.}
	\label{fig6}
\end{figure}

\begin{figure}
\centering
		\includegraphics[width=1 \columnwidth]{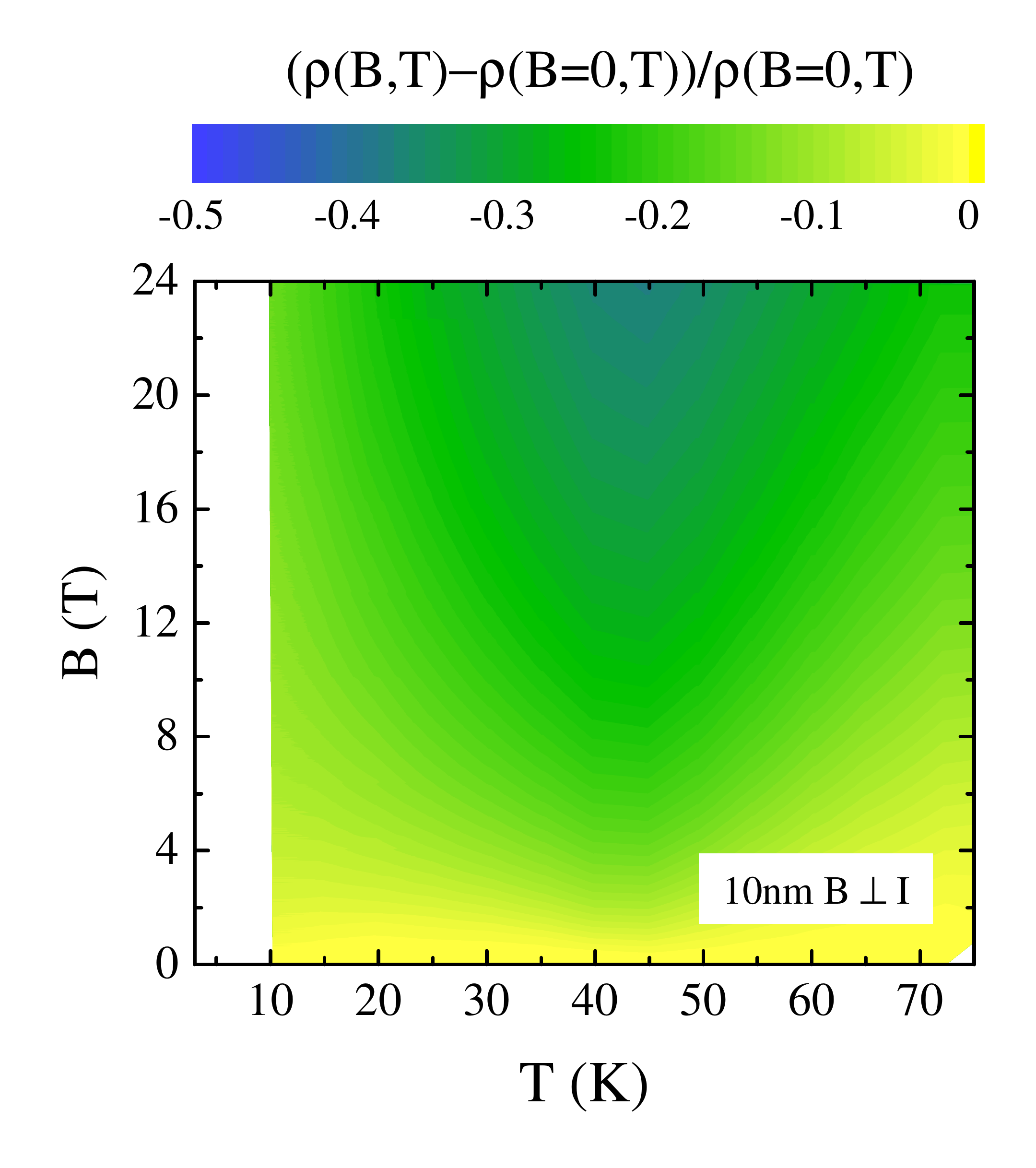}
	\caption{(Color online) Contour plot of the magnetoresistivity $MR$ in transverse magnetic fields up to 24\,T for a 10\,nm thin film MnSi; for details see text.}
	\label{fig7}
\end{figure}

\begin{figure}
\centering
		\includegraphics[width=1 \columnwidth]{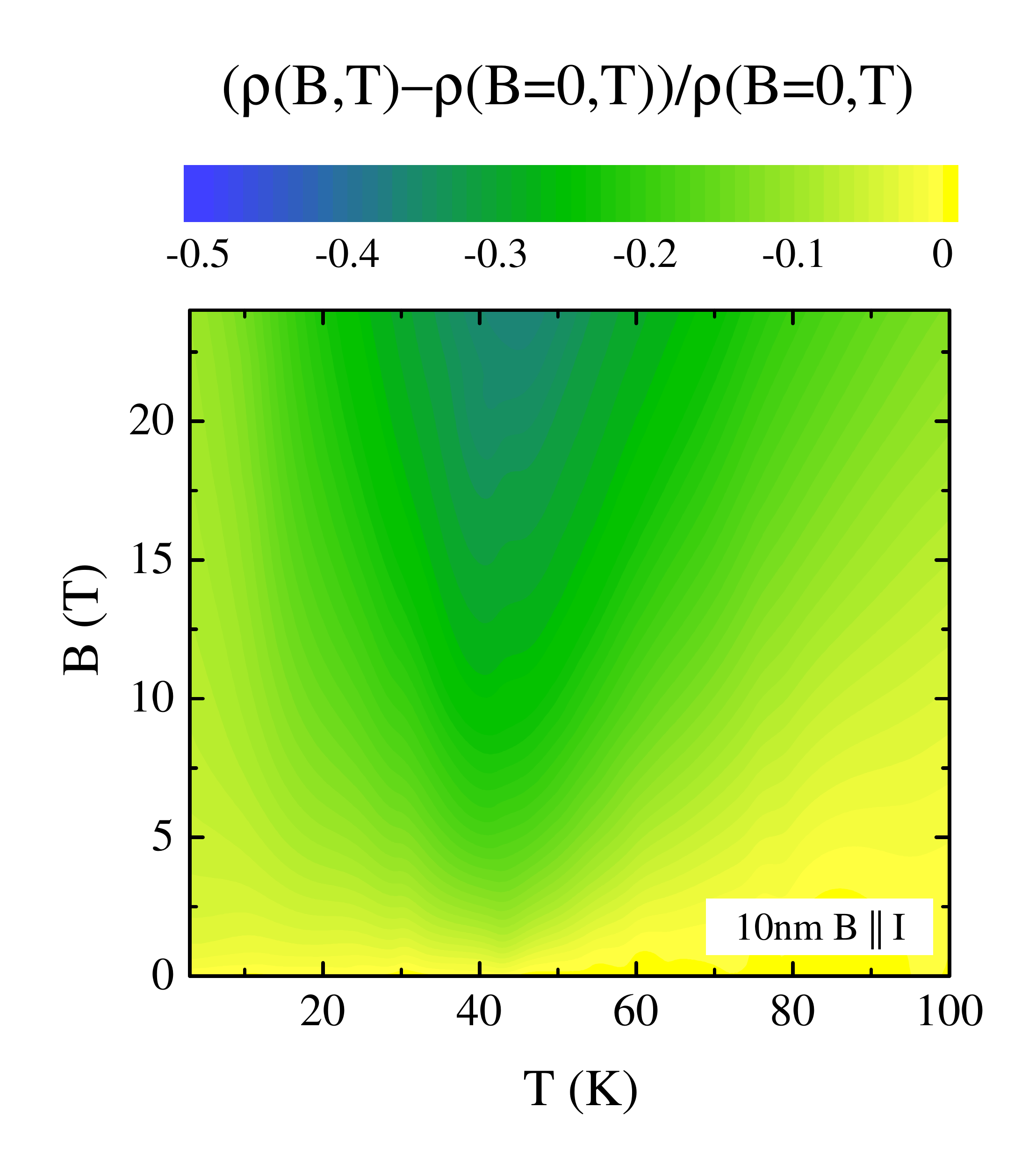}
	\caption{(Color online) Contour plot of the magnetoresistivity $MR$ in parallel magnetic fields up to 24\,T for a 10\,nm thin film MnSi; for details see text.}
	\label{fig8}
\end{figure}

Finally, to illustrate the similar behavior seen for both our samples and both field geometries, in the Figs. \ref{fig6} - \ref{fig8} we plot the contour plots of the magnetoresistivity for the remaining data sets. Overall, there is a close similarity for all data sets, be it that the magnetoresistive effects are somewhat weaker for the 10\,nm sample compared to the 30\,nm film. Likely, it simply reflects the larger zero field resistivity of the 10\,nm sample, which will reduce the overall signal size of the $MR$ (see Fig. \ref{fig1}).

\section{DISCUSSION}

So far, experimentally we have demonstrated two points: a.) There is basically no difference in the magnetoresistive response of films MnSi with different thickness and for different field directions. Thus, structurally the films are still in a 3D-limit, consistent with the argument that size effects induced by the film thickness only occur below $\sim$\,10\,nm thickness \cite{engelke2012}. As well, the negligible difference between longitudinal and transverse magnetoresistivity likely reflects the morphology of our films, {\it i.e.}, epitaxial growth of MnSi islands with a typical diameter of $\sim 50$\,nm \cite{schroeter2018}. A significant scattering contribution will thus arise from the grain boundaries and surfaces, which are present for both experimental geometries. Conversely, effects such as the existence of skyrmions, domains, Fermi surface anisotropies etc. that might lead to a difference of longitudinal and transverse magnetoresistivity will only have a secondary relevance. b.) On a qualitative and semi-quantitative level there is a close resemblance of the magnetoresistivity of thin film MnSi to that of single crystalline material. We thus can proceed and carry out a data analysis as has been put forth by Sakakibara et al. \cite{sakakibara1982}. 

\begin{figure}
\centering
		\includegraphics[width=1 \columnwidth]{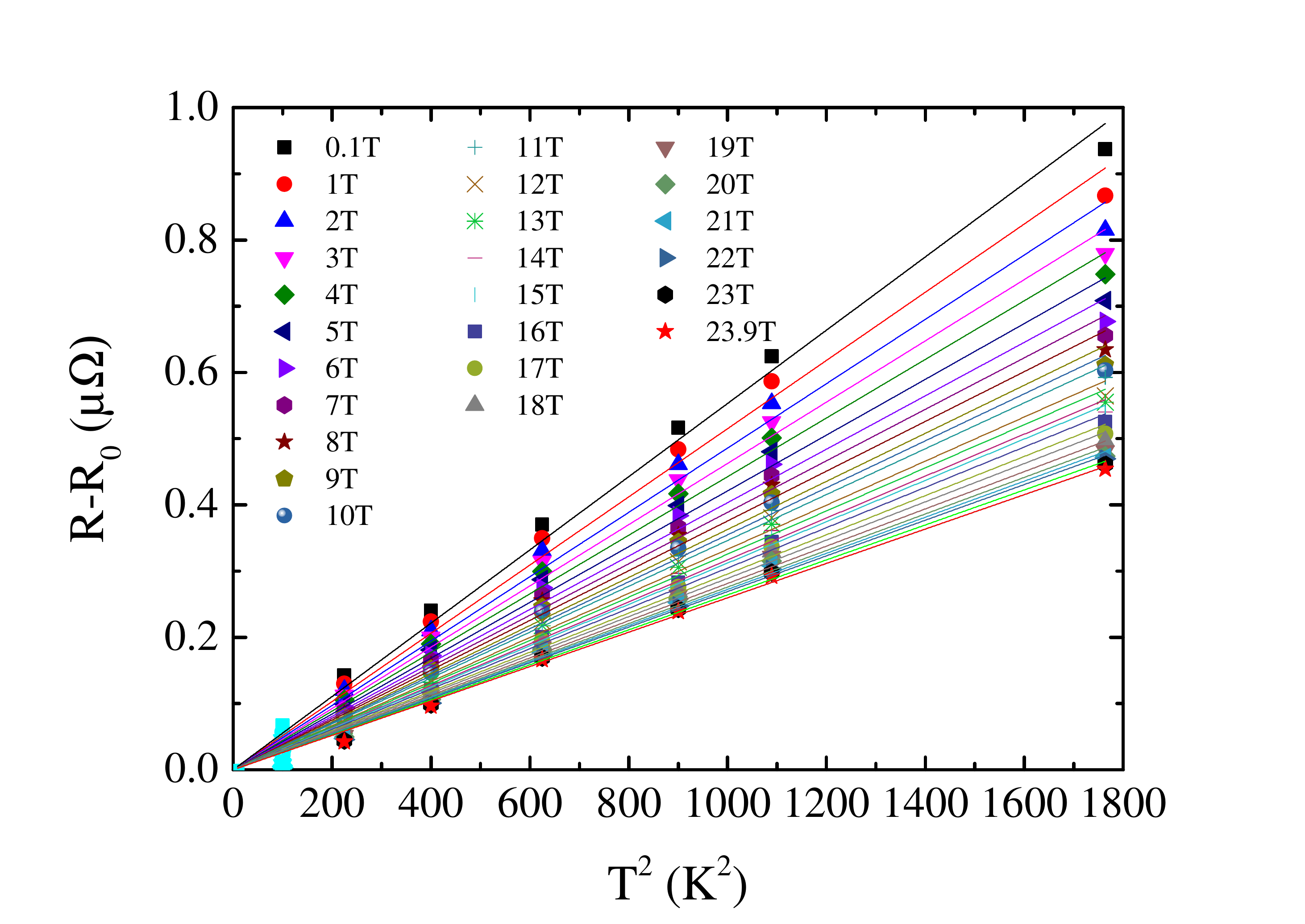}
	\caption{(Color online) Plot for data analysis of the magnetoresistivity $MR$ in parallel magnetic fields up to 24\,T for a 10\,nm thin film MnSi; for details see text.}
	\label{fig9}
\end{figure}

We start by noting that for weakly and nearly ferromagnetic metals the in-field dependence of the low-temperature magnetoresistivity can be expressed as \cite{ueda1976}
\begin{center}
$R(T,B)=R_{0}+R_{2}(B)T^{2}$.
\end{center}
$R_{0}$ is the residual resistance at 0\,K, while the second term reflects the spin fluctuation effect of an itinerant weakly or nearly ferromagnetic material. The factor $R_{2}(B)$ of the spin fluctuation term can be determined from the $MR$ data by plotting $R(T,B)-R_{0}$ over $T^{2}$. For $T < T_{N}$ and in high magnetic fields, in this representation the quantity $R(T,B)-R_{0}$ results in straight lines vs. $T^2$, were the slope corresponds to $R_{2}(B)$. In Fig. \ref{fig9} we plot the experimental data for the 10\,nm film in this representation, verifying that our approach to analyze the data properly reproduces our experimental findings.

With respect to this type of analysis, since the residual resistivity $R_{T,B=0}$ has a temperature dependence because of phonon scattering, strictly speaking we should plot $R(T,B)-R(T,B=0) \propto T^{2}$. In fact, as a first approximation, allowing $R_0$ as a free fitting parameter to vary by about 10 \% slightly improves the fitting, but it does not affect the fundamental outcome of the data analysis, {\it i.e.}, the field dependence $R_{2}(B)$.

\begin{figure}
\centering
		\includegraphics[width=1 \columnwidth]{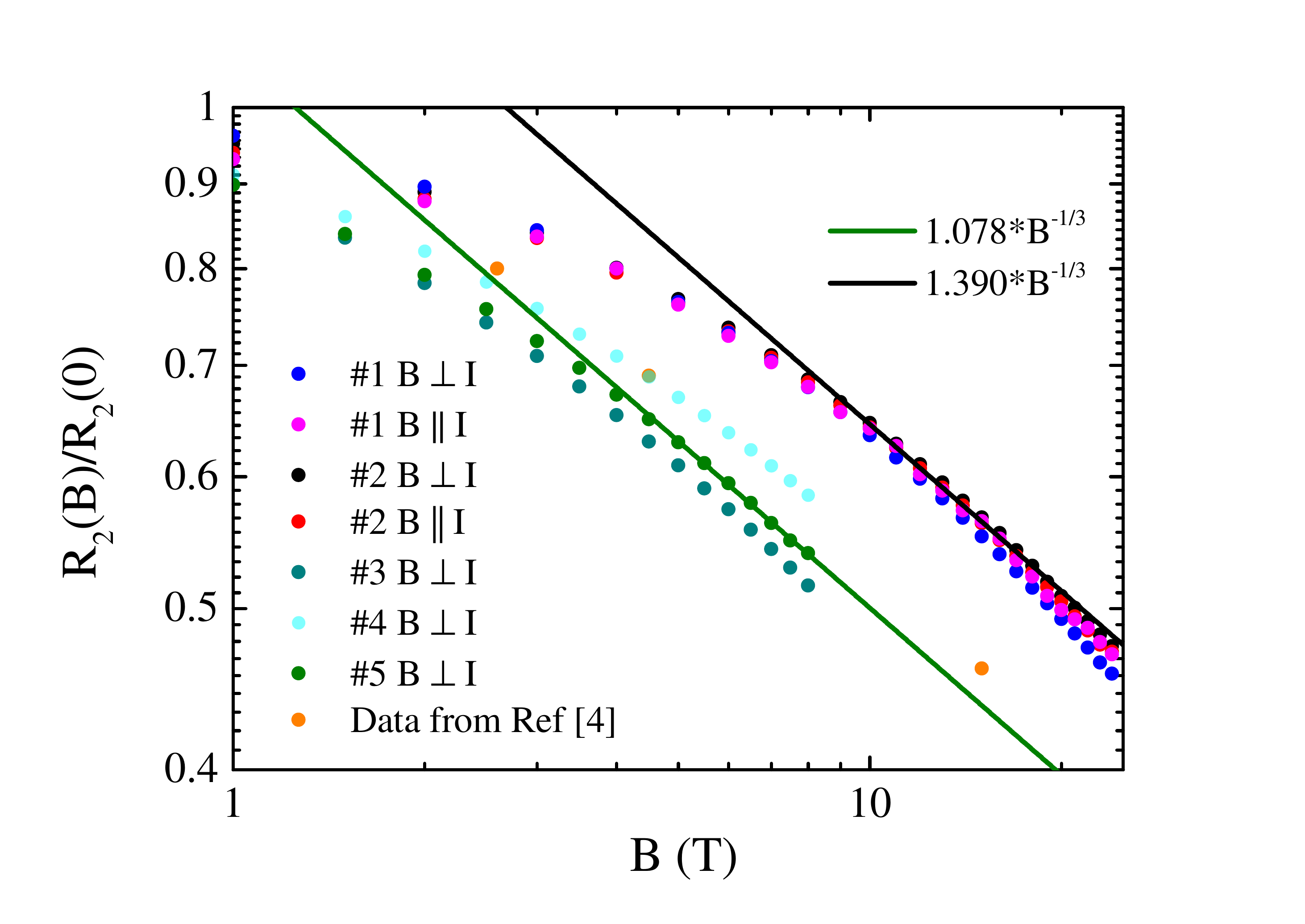}
	\caption{(Color online) Field dependence of the magnetoresistive coefficient $R_{2}(B)/R_{2}(B=0)$ of bulk and thin film MnSi (\#1: thin film 10 nm thickness, \#2: thin film 30 nm thickness, \#3: bulk with RRR of 15, \#4: bulk with RRR of 42, \#5: bulk with RRR of 104); for details see text.}
	\label{fig10}
\end{figure}

Next, we follow the argumentation set out in Ref. \cite{sakakibara1982} and plot the normalized field dependence of the resistive coefficient $R_{2}(B)/R_{2}(B=0)$ in Fig. \ref{fig10}. Here, we include the data from our two thin film samples for both field geometries, from the single crystals measured as references, and the data published in Ref. \cite{sakakibara1982}. By plotting a normalized quantity, we get around the uncertainties in the determination of absolute resistivity values.

From the plot, it is immediately clear that overall thin film and single crystal samples exhibit a qualitatively similar behavior, while quantitatively there are clear differences. First, for single crystalline material, and even under consideration of the limited field range for the crystals measured here, our data sets essentially reproduce those from Sakakibara et al. \cite{sakakibara1982} on single crystalline samples. With the residual resistivity ratios for the crystals used for this plot varying by almost an order of magnitude, we conclude that disorder does not significantly affect the magnetoresistive behavior, although there might be some effect hidden in the data.

Next, it was pointed out that in high fields the magnetoresistive behavior of a weakly or nearly ferromagnetic metal such as MnSi should have a field dependence evolving like $R_{2}(B)/R_{2}(B=0) \propto B^{-1/3}$ \cite{sakakibara1982,ueda1976}. The solid lines in Fig. \ref{fig10} visualize such a field dependence, which to good approximation is fulfilled both for single crystal and thin film MnSi in a similar high field range. Consequently, the magnetoresistive response of thin film MnSi can essentially be understood within what is nowadays labeled the self-consistent renormalization (SCR) theory of spin fluctuations \cite{moriya1985}. Conversely, the quantitative difference between thin films and single crystals must reflect a difference of the microscopic parameters used within SCR theory to describe the spin fluctuations. 

In detail, by writing out the expressions for the high field magnetoresistivity given in Ref. \cite{ueda1976}, one finds that $R_{2}(B)/R_{2}(B=0)$ is a complex function of variables introduced in SCR theory to parametrize the spin fluctuations: $R_{2}(B)/R_{2}(B=0) \propto \sqrt{\frac{\chi_0}{\chi L^{1/3}}} B^{-1/3}$, with $\chi_0$ the susceptibility of the non-interacting system, $\chi$ the susceptibility of the interacting system, and $L$ an expansion coefficient of the magnetic free energy. 

Evidently, we cannot extract unique values for these parameters from our experiment. However, comparing the experimental results for single crystalline and thin film material MnSi, the quantitative differences imply that the spin fluctuation spectrum in our thin films is different from the single crystals. At this point, it is not clear if this just reflects the effective negative pressure in the thin films. Alternatively, the uni-axial anisotropy induced in the films, which was argued to substantially affect film properties \cite{karhu2012,wilson2014}, might cause modifications of the spin fluctuation spectra. One way to test these notions would be - for instance - corresponding magnetoresistivity measurements under pressure on thin film MnSi. If the pressure scenario holds, we would expect a gradual transition of the magnetoresistive behavior of the films towards the bulk behavior with applied pressure. Altogether, while in terms of spin fluctuations thin film and bulk (single crystal) MnSi can be understood within the same SCR theoretical framework, our findings imply that thin films are in a different spin fluctuation parameter range than bulk (single crystalline) material.   

\begin{acknowledgments}
We acknowledge the support of the LNCMI-CNRS, member of the European Magnetic Field Laboratory (EMFL). We gratefully acknowledge support by the Braunschweig International Graduate School of Metrology B-IGSM and the DFG Research Training Group GrK1952/1 ''Metrology for Complex Nanosystems''.
\end{acknowledgments}

\appendix

\end{document}